\documentclass[prb,twocolumn,superscriptaddress,nofootinbib,longbibliography]{revtex4-1}
\usepackage{amsfonts,amssymb,amsmath,latexsym}
\usepackage{graphicx}
\usepackage{xcolor}
\usepackage{setspace}
\usepackage[utf8]{inputenc}
\usepackage{graphicx}
\usepackage{multirow}
\usepackage{tabularx}
\usepackage{soul}
\newcommand{\br}{\vec{r}}

\newcommand{\bn}{\vec{n}}

\newcommand{\bu}{\vec{u}}
\newcommand{\bv}{\vec{v}}

\newcommand{\bR}{\vec{R}}
\newcommand{\bq}{\vec{q}}
\newcommand{\ba}{\vec{a}}
\newcommand{\bA}{\vec{A}}
\newcommand{\bG}{\vec{G}}

\newcommand{\cemes}{CEMES/CNRS, 29 rue J. Marvig, 31055 Toulouse, France}
\newcommand{\lcpq}{Laboratoire de Chimie et Physique Quantiques, CNRS, Universit\'e de Toulouse, UPS, 118 route de Narbonne, F-31062 Toulouse, France}
\newcommand{\unibo}{Universit\`a di Bologna, Bologna, Italy}
\newcommand{\etsf}{European Theoretical Spectroscopy Facility (ETSF)}

\begin{document}
\title{Accurate ground-state energies of Wigner crystals from a simple real-space approach}

\author{Estefania Alves}
\affiliation{\cemes}
\author{Gian Luigi Bendazzoli}
\affiliation{\unibo}
\author{Stefano Evangelisti}
\email{stefano.evangelisti@irsamc.ups-tlse.fr}
\affiliation{\lcpq}
\author{J.~Arjan Berger}
\email{arjan.berger@irsamc.ups-tlse.fr}
\affiliation{\lcpq}
\affiliation{\etsf}

\date{\today}

\begin{abstract}
\label{abstract}
We propose a simple and efficient real-space approach for the calculation of the ground-state energies of Wigner crystals in 1, 2, and 3 dimensions. To be precise, we calculate the first two terms in the asymptotic expansion of the total energy per electron which correspond to the classical energy and the 
harmonic correction due to the zero-point motion of the Wigner crystals, respectively. 
Our approach employs Clifford periodic boundary conditions to simulate the infinite electron gas and a renormalized distance to evaluate the Coulomb potential. This allows us to calculate the energies unambiguously and with a higher precision than those reported in the literature. 
Our results are in agreement with the literature values with the exception of harmonic correction of the 2-dimensional Wigner crystal for which we find a significant difference.
Although we focus on the ground state, i.e., the triangular lattice and the body-centered cubic lattice, in two and three dimensions, respectively, we also report the classical energies of several other common lattice structures.
\end{abstract}

\maketitle

%
\section{Introduction}
The uniform electron gas (UEG)~\cite{GiulianiVignale,Loos_2016}, otherwise known as jellium, has proven to be a very useful model for the understanding of electron interactions. In particular in the solid state the UEG can be used to study a variety of phenomena, such as plasmon oscillations~\cite{Tonks_1929}, electron screening~\cite{Lindhard_1954}, the quantum Hall effect~\cite{Klitzing_1980} and Wigner localization~\cite{Wigner_1934,Jauregui_1993,Diaz-Marquez_2018,Escobar_2019,Ziani_2020}.
Moreover, by combining the UEG with density-functional theory (DFT) predictive calculations can be performed on both solids and molecules.
Thanks to quantum Monte Carlo calculations~\cite{Ceperley_1980} the correlation contribution to the ground-state energy of the UEG as a function of the density is accurately known. This data has been used to approximate the unknown exchange-correlation energy of DFT.~\cite{Vosko_1980, Perdew_1981,Perdew_1992}

Almost a century ago, it was predicted by Wigner~\cite{Wigner_1934} that in the limit of an infinitely dilute UEG the electrons crystallize at fixed positions in space, thus forming a crystal lattice.
Wigner crystals have later been experimentally realized in 1D and 2D and have shown to exhibit interesting properties.~\cite{Shapir_2019,Grimes_1979}
However, numerical calculations were required to determine the ground-state crystal structures in both 2D and 3D. By comparing the energies of several Bravais lattices it was concluded that in 2D the electrons crystallize in the triangular structure~\cite{Bonsall_1977} while in 3D they crystallize in the body-centered cubic structure.~\cite{Wigner_1934,Fuchs_1935,Sholl_1967,Hasse_1991}

The energy per electron of a Wigner crystal $E_{WC}$ can be written as an asymptotic expansion in powers of $r_s^{-\frac{1}{2}}$:~~\cite{Carr_1961_Jun}
\begin{equation}
\label{expansion}
E_{WC}\sim\frac{\eta_{0}}{r_s}+\frac{\eta_{1}}{r_s^{3/2}}+\frac{\eta_{2}}{r_s^2}+\frac{\eta_{3}}{r_s^{5/2}}+... \,\,,
\end{equation}
where $r_s$ is the Wigner-Seitz radius. The first term on the right-hand side is the energy corresponding to a classical charge distribution, 
the second term is a correction due to the zero-point motion in the harmonic approximation, while $\eta_2, \eta_3, \cdots$ correspond to anharmonic corrections. 
In this work we will focus on the calculation of $\eta_0$ and $\eta_1$.
These parameters have been calculated in the past using reciprocal-space approaches.
Here we will show that they can also be calculated within a simple real-space method.
We note that Eq.~\eqref{expansion} assumes that the electrons are discernable.~\cite{Wigner_1934}
However, for large $r_s$ the corresponding error is negligible since the correction falls off exponentially with $r_s$.

For the 3D Wigner crystal the first accurate calculation of the classical ground-state energy per electron was done by Fuchs.~\cite{Fuchs_1935} 
He obtained $\eta_0^{\text{3D}} = -0.895\,93 \,\, \text{Ha}$. This value was later refined to $\eta_0^{\text{3D}}= -0.895\,929 \,\, \text{Ha}$ by Hasse and Avilov ~\cite{Hasse_1991}.
The first estimation of the harmonic correction was done by Wigner who found $\eta_1^{\text{3D}} = 2.7$ Ry.~\cite{Wigner_1938}
This result is quite close to those obtained two decades later by Coldwell-Horsfall and Maradudin and by Carr.
They found $\eta_1^{\text{3D}} = 1.319 \,\, \text{Ha}$~\cite{Coldwell_1960} and $\eta_1^{\text{3D}} = 1.33 \,\, \text{Ha}$,~\cite{Carr_1961_Jun} respectively.
 Using the same approach as Carr but with an improved integration over the Brilliouin zone, Nagai and Fukuyama found the most accurate value to date, i.e., $\eta_1^{\text{3D}} = 1.328\,62 \,\, \text{Ha}$.~\cite{Nagai_1982} 
 
Both the classical ground-state energy per electron and the harmonic correction of the 2D Wigner crystal have been calculated by Bonsall and Maradudin~\cite{Bonsall_1977}. They found $\eta_0^{\text{2D}} = 1.106\, 103 \,\, \text{Ha}$ and $\eta_1^{\text{2D}} = 0.795 \,\, \text{Ha}$.
To the best of our knowledge this is the only calculation for $\eta_1^{\text{2D}}$ in the literature.
Below we will show that we obtain a value with a much higher precision and which differs significantly from the value of Bonsall and Maradudin.
 
In 1D the classical ground-state energy per electron diverges because of a non-integrable singularity in the 1D Coulomb potential in the origin. Therefore, we will not consider the calculation of $\eta_0^{\text{1D}}$. We note, however, that regularization techniques can be used for the Coulomb potential to allow for its calculation.~\cite{Fogler_2005,Loos_2013,Loos_2014}. Instead, $\eta_1^{1D}$ can be calculated without problems and its value has been determined with high accuracy.~\cite{Fogler_2005,Loos_2013,Loos_2014}
Its 6-digit approximation is $\eta_1^{1D} = 0.359\,933 \,\, \text{Ha}$.
In table \ref{Table:literature} we summarize the most accurate values for $\eta_0$ and $\eta_1$ that can be found in the literature.
\begin{table}
\caption{Summary of the most accurate literature values (in Ha) of the coefficients $\eta_0$ and $\eta_1$.}
\begin{center}
\begin{tabular}{lccc}
\hline
\hline
 & 1D &  2D &  3D   \\
 & \, linear lattice \,& \, triangular lattice \, & \, bcc lattice \,  \\ 
\hline
$\eta_0$ &  - & -1.106 103~\cite{Bonsall_1977}  & -0.895 929~\cite{Hasse_1991}    \\
$\eta_1$ &  0.359 933~\cite{Fogler_2005} & 0.795~\cite{Bonsall_1977} &   1.328 62~\cite{Nagai_1982}   \\
\hline
\end{tabular}
\end{center}
\label{Table:literature}
\end{table}

The goal of this work is two-fold: 1) to present a simple and general real-space approach for the calculation of the coefficients $\eta_0$ and $\eta_1$ and 2) to give a  larger precision of those coefficients, in particular for $\eta_1$, for Wigner crystals in 2D, and 3D.
We will use an approach based on Clifford boundary conditions and a renormalized distance~\cite{Valenca_2019} that we previously have successfully applied to the calculation of Madelung constants.~\cite{Tavernier_2020,Tavernier_2021}

The paper is organized as follows. In section \ref{Sec:theory} we describe the theoretical details of our real-space approach.
In section \ref{Sec:results} we discuss our results for the energies of the Wigner crystals.
Finally, in section \ref{Sec:conclusions} we draw our conclusions.
We use Hartree atomic units $(\hbar = e= m_e = a_0 = 1)$ throughout this work.
\section{Theory}
\label{Sec:theory}
\subsection{The jellium model}
The Hamiltonian of an infinite uniform electron gas with a uniform positive background that ensures the charge neutrality of the system is given by
\begin{equation}
\label{Hamiltonian}
\hat{H}_{jellium} = -\sum_i \frac{\nabla_{\vec{r_{i}}}^2}{2} + \hat{U}_{ee} + \hat{U}_{bb} + \hat{U}_{eb},
\end{equation}
in which the electron-electron, electron-background, and background-background contributions to the Coulomb potential are given by, respectively,
\begin{align}
\hat{U}_{ee} & =  \frac{1}{2}\sum_{\substack{i,j\\i \neq j}}\frac{1}{|\vec{r}_{i}-\vec{r}_{j}|},
\\
\hat{U}_{eb} & = -\sum_{i}\int d\br\frac{n}{| \br-\vec{r}_{i}|},
\\
\hat{U}_{bb} & = \frac{1}{2}\int d\br \int d\br^{\,\prime} \frac{n^2}{|\br - \br^{\,\prime}|},
\end{align}
where 
$n$ is the uniform positive background density.
The charge neutrality of the system is imposed by assuming that the constant electron density is equal to the positive background density $n$.
Individually each term in Eq.~\eqref{Hamiltonian} diverges but their sum is finite, except for the one-dimensional uniform electron gas.

At low density the electrons form a Wigner crystal with the electrons localized at the lattice positions of a crystal.
Therefore, we can perform a Taylor expansion of the Coulomb potential around the equilibrium lattice vectors $\bR$ of the electrons in the Wigner crystal,
\begin{equation}
\label{Hamiltonian_exp}
\hat{H}_{jellium} = -\frac12 \sum_i \nabla_{\vec{r_{i}}}^2 + \hat{U}_0 + \hat{U}_2 
+ \cdots \,\,  ,
\end{equation}
in which
\begin{align}
\hat{U}_0 &=\frac{1}{2}\sum_{\substack{i,j\\i \neq j}}\frac{1}{|\bR_{i}-\bR_{j}|} -\sum_{i}\int d\br\frac{n}{|\br-\bR_{i}|} + \hat{U}_{bb}
\label{Eqn:U0}
\\
\notag
\hat{U}_2 &=\frac12 \sum_{mn} \sum_{\alpha\beta}
\left. \partial_{m\alpha}\partial_{n\beta} \hat{U}_{ee}\right|_{{\br}_{x}=\bR_{x}\forall x} 
\\ & \times
({r}_{m,\alpha}  - R_{m,\alpha}) ({r}_{n,\beta} - R_{n,\beta}),
\label{Eqn:U2}
\end{align}
where the Greek letters $\alpha$ and $\beta$ denote Cartesian components.
Since the classical energy $U_0$ is a minimum, the contribution of the first order term in the expansion, $\hat{U}_1$, vanishes.
Furthermore, only variations in $\hat{U}_{ee}$ contribute to $\hat{U}_2$ since $\hat{U}_{bb}$ is independent of the electronic coordinates and variations in the electronic position do not change $\hat{U}_{be}$ because of the uniformity of the background.
In this work we will consider the first three terms on the right-hand side of Eq.~\eqref{Hamiltonian_exp}.
This allows us to calculate the first two coefficients in Eq.~\eqref{expansion}.

Defining the relative coordinates $\bu_m = \br_m - \bR_m$ 
we can rewrite the first three terms of Eq.~\eqref{Hamiltonian_exp} in the following general form~\cite{Carr_1961_Jun} 
\begin{equation}
\hat{H} =
\label{Eqn:H_trunc}
\hat{U}_0 - \frac12 \sum_i \nabla_{\bu_i}^2 +
\frac12 \sum_{mn} \sum_{\alpha\beta}
C_{\alpha\beta}(\bR_m - \bR_n) 
u_{m,\alpha}u_{n,\beta},
\end{equation}
in which the real symmetric matrix $\mathbf{C}$ is defined as
\begin{equation}
\label{Eqn:C}
C_{\alpha\beta}(\bR_m - \bR_n) =
\frac12 \partial_{m\alpha}\partial_{n\beta}  \sum_{\substack{i,j\\i \neq j}} \left. \frac{1}{|\bu_i - \bu_j +\bR_i - \bR_j|} \right|_{{\bu}_x=0\forall x}
\end{equation}
where the derivatives are now with respect to $\bu$.

Thanks to the translational invariance of the system we can use the following Fourier transformation
\begin{align}
\bu_{m} &= \frac{1}{\sqrt{\mathcal{N}}}\sum_{k} e^{i\bG_k\cdot\bR_m} \bv_{k}
\\
\bv_{k} &= \frac{1}{\sqrt{\mathcal{N}}}\sum_{m} e^{-i\bG_k\cdot\bR_m} \bu_{m},
\end{align}
where $\mathcal{N}$ is a normalisation constant and the vectors $\bG_k$ can be interpreted as reciprocal lattice vectors,
to rewrite Eq.~\eqref{Eqn:H_trunc} as
\begin{equation}
\hat{H} = \hat{U}_0 +
\sum_{k} \left[
-\frac12 \nabla_{\bv_k}^2  + \frac12\sum_{\alpha\beta}\tilde{C}_{\alpha\beta}(\bG_k) v_{k,\alpha} v_{k,\beta}^*
\right]
\end{equation}
with
\begin{equation}
\label{Eqn:Ctilde}
\tilde{C}_{\alpha\beta}(\bG_k) = \sum_l C_{\alpha\beta} (\bR_l) e^{i\bG_k \cdot \bR_l}
\end{equation}
a real symmetric $d\times d$ matrix with $d=1,2,3$ the dimensionality of the lattice.
Finally, using the eigenvectors of $C (\bG_k)$ we can perform an orthonormal transformation to arrive at
\begin{equation}
\hat{H}= \hat{U}_0 + 
\sum_{k}\sum_{\alpha} \left[
-\frac12\partial_{k,\alpha}^2  +  \frac12\omega_{k,\alpha}^2 q^2_{k,\alpha}
\right],
\end{equation}
where $\omega_{k,\alpha}^2$ are the eigenvalues of $\tilde{C}(\bG_k)$ and $\bq_k$ are normal modes.
The expression inside the square brackets in the above equation is the Hamiltonian of a quantum harmonic oscillator in one dimension for which the eigenenergies are known.
Therefore, the total ground-state energy per electron of the Wigner crystal can be written as
\begin{equation}
E_{WC} \sim \frac{\eta_0}{r_s} + \frac{\eta_1}{r_s^{3/2}} + \cdots,
\end{equation}
with
\begin{align}
\label{Eqn:eta0}
\frac{\eta_0}{r_s} & = \frac{U_0}{N}
\\
\label{Eqn:eta1}
\frac{\eta_1}{r_s^{3/2}} & = \frac{1}{2N}\sum_{k} \sum_{\alpha} \omega_{k,\alpha}.
\end{align}
The jellium problem pertains to a system with an infinite number of electrons in an infinite volume at constant density.
In practical calculations we can of course only deal with a finite number of electrons in a finite volume.
However, we would like to preserve the translational symmetry of the jellium model.
Therefore, we impose periodic boundary conditions with respect to a finite supercell.
Unfortunately, the long-range Coulomb potential is not periodic and it does not vanish at the borders of any, even very large, finite supercell.
Therefore, as explained in the next section, we impose Clifford boundary conditions with a renormalized distance.
\subsection{Clifford boundary conditions}
We will use Clifford boundary conditions which means that we define a supercell that has the topology of a Clifford torus, i.e., a finite, flat and borderless manifold.
A Clifford supercell is linked to a Euclidean supercell defined in $\mathbb{R}^d$.
The Clifford supercell is then obtained by joining opposite sides of the Euclidean supercell \emph{without} deformation.
This can be achieved by defining the Clifford supercell in the embedding space $\mathbb{C}^d$ (alternatively, it can also be achieved in $\mathbb{R}^{2d}$).

Let us consider a general orthorhombic lattice in $d$ dimensions and define $\ba_{\alpha}$ 
to be the orthogonal generating vectors of a unit cell in $\mathbb{R}^d$.
Then a general vector $\bv$  inside the unit cell can be written as
\begin{equation}
\bv = \sum_{\alpha} x_{\alpha} \ba_{\alpha},
\end{equation}
where $0 \le x_{\alpha} < 1$.
We define a Euclidean supercell (ESC) as the right parallelepiped in $\mathbb{R}^d$ generated by the vectors $\bA_{\alpha}$ given by
\begin{equation}
\bA_{\alpha} = K_{\alpha} \ba_{\alpha}.
\end{equation}
where $K_{\alpha}$ are positive integers.
The ESC thus contains $\prod_{\alpha} K_{\alpha}$ copies of the unit cell.
A general vector $\vec{r}^{\,\,ESC}$ in the ESC can thus be written as
\begin{equation}
\vec{r}^{\,\,ESC} = \bv + \sum_{\alpha} k_{\alpha} \ba_{\alpha} = \sum_{\alpha} r_{\alpha} \ba_{\alpha}
\end{equation}
with $r_{\alpha} = x_{\alpha} + k_{\alpha}$ and $0\le k_{\alpha} \le K_{\alpha} -1$.

We now define the Clifford supercell (CSC) as the Clifford torus in which the opposite borders (either points, edges, or faces, depending on $d$) of the corresponding ESC are connected without deformation.
A general vector $\vec{r}^{\,\,CSC}$ in the CSC should respect the translational symmetry of the Clifford torus.
This can be achieved by writing $\vec{r}^{\,\,CSC}$ according to
\begin{equation}
\vec{r}^{\,\,CSC} = \sum_{\alpha}
\frac{K_{\alpha}}{2\pi \mathrm{i}} \left[e^{\mathrm{i} 2\pi r_{\alpha} / K_{\alpha}} -1\right] \ba_{\alpha}.
\end{equation}
We note that the above expression is the classical equivalent of the PBC position operator proposed by some of us in a quantum context.~\cite{Valenca_2019}
The above definition satisifies a number of important constraints.
In particular, it satisfies the translational symmetry, it reduces to the standard position operator in the appropriate limit and the corresponding definition of the distance is real and gauge invariant (see Eq.~(\ref{Eqn:dist_renorm2}) below).~\cite{Valenca_2019}

\begin{figure}[t]
\includegraphics[width=\columnwidth]{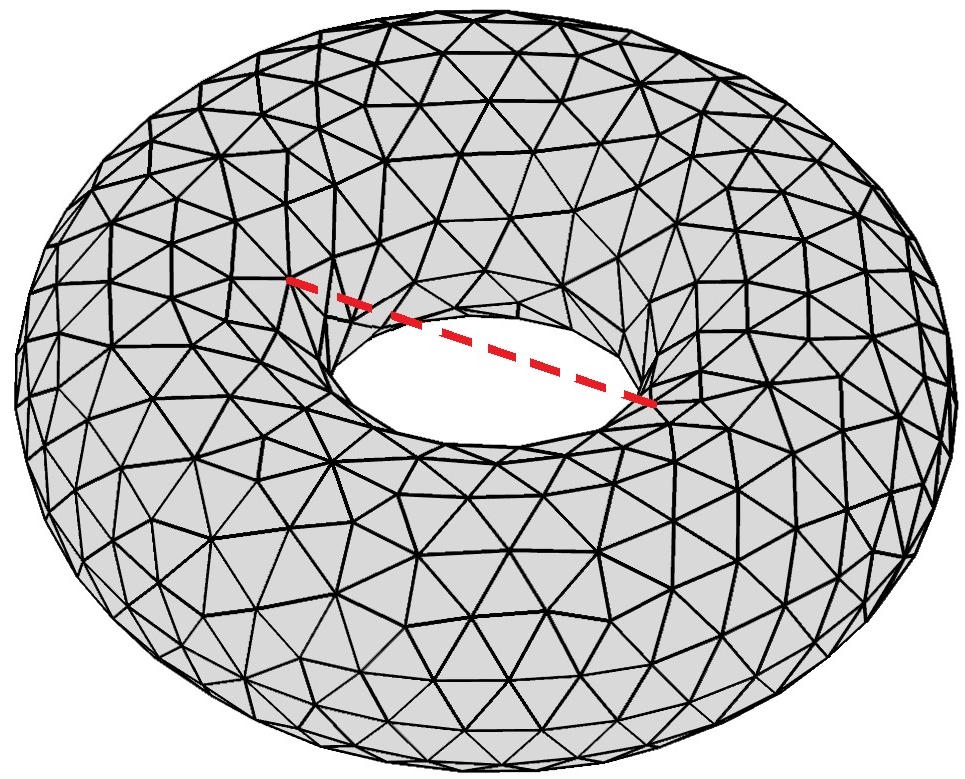}
\caption{Pictorial illustration of a Clifford supercell for the triangular lattice of the 2-dimensional Wigner crystal.
The equilibrium positions of the electrons are located at the vertices.
The dashed red line represents the renormalized distance between two electrons used in the Coulomb potential. 
It is the shortest distance in the embedding space of the Clifford torus. 
We note that a true Clifford torus has a flat surface which is impossible to represent graphically.}
\label{Fig:torus}
\end{figure}

In order to evaluate Coulomb potentials we have to define the distance between two points in the CSC.
We note that a possibility would be to define the distance as the smallest difference between two points \emph{on} the Clifford torus.
However, such a distance would have discontinuous derivatives in those points of the CSC that correspond to the midpoints of the edges of the corresponding ESC.
This would yield discontinuous forces in these points, which is unphysical.
Therefore, we choose the distance to be the Euclidean norm in $\mathbb{C}^d$ because it is both uniquely defined and yields continuous derivatives.
In other words, the distance is defined in the embedding space of the Clifford torus.
This distance $r^{CSC}_{ij} = |\vec{r}^{\,\,CSC}_i - \vec{r}^{\,\,CSC}_j|$ is given by
\begin{equation}
r^{CSC}_{ij} = 
\sqrt{\sum_{\alpha} 
\frac{L_{\alpha}^2}{\pi^2}\sin^2\left(\frac{\pi}{L_{\alpha}}[r_{i\alpha} - r_{j\alpha}]\right)},
\label{Eqn:dist_renorm2}
\end{equation}
where we used that $\ba_{\alpha} \cdot \ba_{\beta} = 0$ for $\alpha \neq \beta$ and $L_{\alpha} = K_{\alpha} |{\ba}_{\alpha}|$ with $L_{\alpha}$ the length of an edge of the ESC.
We will evaluate the Coulomb potentials in Eqs.~\eqref{Eqn:U0} and \eqref{Eqn:U2} using the above renormalized distance.
In Fig.~\ref{Fig:torus} we show an illustration of a CSC for a 2-dimensional Wigner crystal and the renormalized distance between the electrons.
\subsection{The 3D Wigner crystal}
With the renormalized distance the background-background contribution in the 3D CSC is given by
\begin{equation}
\hat{U}_{bb}^{CSC} \!\! = \frac{N^2}{2V}\int_V
\frac{dx\, dy\, dz}{\sqrt{\frac{L_x^2}{\pi^2}\sin^2[\frac{\pi x}{L_x}]+\frac{L_y^2}{\pi^2}\sin^2[\frac{\pi y}{L_y}]+\frac{L_z^2}{\pi^2}\sin^2[\frac{\pi z}{L_z}]}}
\label{Eqn:U_bb_3D}
\end{equation}
where, thanks to the periodicity of the CSC, we could reduce the two volume integrals to only one
and we used that $n = N/V$, which ensures the charge neutrality of the 3D CSC supercell, with $V=L_xL_yL_z$ the volume of the supercell.
With the changes of variable $\theta_x = \frac{\pi x}{L}$, $\theta_y = \frac{\pi y}{L}$, and $\theta_z = \frac{\pi z}{L}$ the above expression
can be rewritten according to
\begin{equation}
\hat{U}_{bb}^{CSC} \!\! =
\frac{N^2}{2\pi^2}\!\!
\int_0^{\pi}\!\!\!\int_0^{\pi}\!\!\!\int_0^{\pi} \!\!\!\! \frac{d\theta_x d\theta_yd\theta_z}
{\sqrt{L_x^2\sin^2{\theta_x}+L_y^2\sin^2{\theta_y}+L_z^2\sin^2{\theta_z}}}.
\end{equation}

The triple integral in the above equation can be readily calculated.
For example, in the case of a cubic supercell, i.e., $L = L_x = L_y = L_z$, we obtain the following result 
\begin{equation}
\hat{U}_{bb}^{CSC} =
\frac{N^2\gamma_c}{L} \quad\quad\quad\quad\quad (L = L_x = L_y = L_z)
\end{equation}
with $\gamma_c = 1.430 505 5275$. 

Thanks to the periodicity and uniformity of the positive background, each electron contributes exactly the same amount to the summation in the expression of the electron-background contribution in the CSC.
Therefore, without loss of generality, we can choose to consider explicitly only the contribution of an electron located at the origin and multiply with $N$.
We can thus write the electron-background contribution according to
\begin{equation}
\hat{U}_{eb}^{CSC}  \!\! = -\frac{N^2}{V}\!\!\int_V \!\!
\frac{dx\, dy\, dz}{\sqrt{\frac{L_x^2}{\pi^2}\sin^2[\frac{\pi x}{L_x}] + \frac{L_y^2}{\pi^2}\sin^2[\frac{\pi y}{L_y}]+ \frac{L_x^2}{\pi^2}\sin^2[\frac{\pi z}{L_z}]}}.
\label{Eqn:U_eb_2D}
\end{equation}
where we once more used $n = N/V$.
We note that in the special case $L = L_x = L_y = L_z$ the integral in the above equation can be made independent of $L$ in a similar way as was done for $\hat{U}_{bb}$.
By comparing Eqs.~\eqref{Eqn:U_bb_3D} and \eqref{Eqn:U_eb_2D} we find the following identity between $\hat{U}_{bb}$ and $\hat{U}_{eb}$ in the CSC,
\begin{equation}
\hat{U}_{eb}^{CSC} = -2\hat{U}_{bb}^{CSC}.
\end{equation}
Finally, the classical electron-electron contribution $\hat{U}_{0,ee}$ in the CSC is given by
\begin{widetext}
\begin{equation}
\hat{U}_{0,ee}^{CSC} =  \frac12\sum_{\substack{i,j\\i \neq j}}\frac{1}{
\sqrt{\frac{L_x^2}{\pi^2}\sin^2[\frac{\pi}{L_x}(x_i - x_j)]+\frac{L_y^2}{\pi^2}\sin^2[\frac{\pi}{L_y}(y_i - y_j)] +\frac{L_z^2}{\pi^2}\sin^2[\frac{\pi}{L_z}(z_i - z_j)]}}
\label{Eqn:Uee_3D}
\end{equation}
\end{widetext}
This is the only contribution that depends on the details of the lattice structure, i.e., the equilibrium positions of the electrons.

In the following we will focus on the body-centered cubic (bcc) lattice since it yields the ground-state energy of a 3D Wigner crystal.
A similar strategy as described below can be used for other lattices.
For the bcc structure it is convenient to use a cubic supercell, i.e., $L=L_x =L_y = L_z$ and to define the equilibrium positions of the electrons according to
$\bR_i = (\pi/3)^{1/3} r_s \bn_i$ with $\bn_i$ a vector of three integers, all even or all odd.~\cite{Carr_1961_Jun} 
Therefore, $L=(\pi/3)^{1/3} (2N_s) r_s$ with $N_s$ the number of electrons per side. 
The total classical bcc energy $U_0 = U_{0,ee}^{CSC} - U_{bb}^{CSC}$ per electron can thus be written as
\begin{equation}
\frac{U_0}{N} =  \frac{\left(\frac{3}{\pi}\right)^{1/3}}{ 2 N_s r_s} 
\left[
\frac{\pi}{2}\sum_{i=1}^N\left(\sum_{\alpha} \sin^2 \left[\frac{\pi n_{i\alpha}}{2N_s}\right] \right)^{-1/2} 
\!\!\!\!-
\gamma_c N \right].
\end{equation}
Since $U_0/N = \eta_0^{bcc}/r_s$ we can easily obtain $\eta_0^{bcc}$ from the above expression.

By working out the derivatives in Eq.~\eqref{Eqn:C} while using the renormalized distance given in Eq.~\eqref{Eqn:dist_renorm2} and then inserting
the definition of $\bR_i$ for the bcc lattice, i.e., $\bR_i = (\pi/3)^{1/3} r_s \bn_i$, we obtain the following expression for the $\mathbf{C}$ matrix
\begin{align}
C^{N_s}_{\alpha\beta}(\vec{0})  = &
\notag
-\frac{3\pi^2}{8N_s^3} \frac{\delta_{\alpha\beta}}{r_s^3} 
\sum_{\bn_i \neq \vec{0}}
\Bigg[
\frac{\cos[\frac{\pi n_{i\alpha}}{N_s}] }{{(\sum_{\alpha'} \sin^2 [\pi n_{i\alpha'}/(2N_s)])^{3/2}}}
\\ &-
\frac{\frac34 \sin^2[\frac{\pi n_{i\alpha}}{N_s}] }
{{(\sum_{\alpha'} \sin^2 [\pi n_{i\alpha'} /(2N_s)])^{5/2}}}\Bigg]
\label{Eqn:C_3D_0}
\\
C^{N_s}_{\alpha\beta}(\bn_i)  = &
\notag
\frac{3\pi^2}{8N_s^3}\frac{1}{r_s^3}
\Bigg[
\frac{\delta_{\alpha\beta}\cos[\frac{\pi n_{i\alpha}}{N_s}]}{{\sum_{\alpha'} (\sin^2 [\pi n_{i\alpha'}/(2N_s)])^{3/2}}}
\\ & -
\frac{\frac34 \sin[\frac{\pi n_{i\alpha}}{N_s}]\sin[\frac{\pi n_{i\beta}}{N_s}]}
{{\sum_{\alpha'} (\sin^2 [\pi n_{i\alpha'}/(2N_s)])^{5/2}}}
\Bigg]
 \quad (\bn_i \neq \vec{0}),
\end{align}
where the summation in Eq.~\eqref{Eqn:C_3D_0} is over all $\bn_i \neq \vec{0}$ inside the CSC.
We note that in the limit $N_s\rightarrow\infty$ we have the following identity
\begin{equation}
\lim_{N_s\rightarrow\infty}  C^{N_s}_{\alpha\alpha}(\vec{0}) = 
\frac{1}{8 r_s^3} .
\end{equation}
The matrix $\tilde{\mathbf{C}}$ given in Eq.~(\ref{Eqn:Ctilde}) can be rewritten as
\begin{equation}
\tilde{\mathbf{C}}^{N_s}(\bn_k) = \sum_{\bn_i} {\mathbf{C}}^{N_s} (\bn_i) \cos\left(\frac{\pi \bn_k \cdot \bn_i}{N_s}\right)
\end{equation}
where we used that $\bG_k =2 \pi \bn_k / (2N_s)$.
To obtain $\eta^{bcc}_1$ it suffices to diagonalize $\tilde{\mathbf{C}}^{N_s}(\bn_k) \, \forall \,  k$, take the square root of the eigenvalues and add them up according to Eq.~(\ref{Eqn:eta1}).
\subsection{The 2D Wigner crystal}
The derivation of the various terms of the Coulomb potential in 2D for the CSC are analogous to those of the 3D Wigner crystal discussed in the previous subsection.
The background-background and electron-background contributions in the CSC are given by
\begin{equation}
\hat{U}^{CSC}_{bb} = -\frac{\hat{U}_{eb}}{2} = 
\frac{N^2}{2\pi}\int_0^{\pi}\int_0^{\pi}\frac{d\theta_x d\theta_y}{\sqrt{L_x^2\sin^2 \theta_x+L_y^2 \sin^2\theta_y}}
\end{equation}
where we used that in 2D $n = N/(L_x L_y)$.
We note that in the special case $L=L_x=L_y$ we obtain the following analytical expression
\begin{equation}
\hat{U}^{CSC}_{bb} = -\frac{\hat{U}_{eb}}{2} = 
\frac{N^2}{2\sqrt{\pi}L} 
G_{3,3}^{2,2} \left(\begin{matrix} 1,1,1 \\
1/2,1/2,1/2
\end{matrix} \middle| 1 \right)
\label{Eqn:U_eb_Meijer}
\end{equation}
where $G$ is the Meijer $G$ function.
The classical electron-electron contribution in the 2D CSC is given by
\begin{equation}
\hat{U}^{CSC}_{0,ee}  =  \frac12\sum_{\substack{i,j\\i \neq j}}\frac{1}
{\sqrt{\frac{L_x^2}{\pi^2}\sin^2[\frac{\pi}{L_x}(x_i - x_j)]+\frac{L_y^2}{\pi^2}\sin^2[\frac{\pi}{L_y}(y_i - y_j)]}}.
\end{equation}

In the following we will focus on the triangular lattice since it yields the ground-state energy of a 2D Wigner crystal.
For the triangular lattice it is convenient to use a rectangular supercell and to define the equilibrium positions of the electrons according to
$\bR_i = r_s \sqrt{\pi/(2\sqrt{3})} (n_{ix}, n_{iy} \sqrt{3})^T$ with $n_x$ and $n_y$ two integers, both even or both odd.
Therefore, $L_x=r_s\sqrt{2\pi/\sqrt{3}} N_s $  and $L_y=r_s \sqrt{2\pi \sqrt{3}}N_s$.
The total classical energy $U_0 = U_{0,ee}^{CSC} - U_{bb}^{CSC}$ per electron of the triangular lattice can thus be written as
\begin{equation}
\frac{U_0}{N} \! =  \!
\frac{1}{N_s r_s} \!\!
\left[ \!
 \frac{\sqrt{\pi}}{2\sqrt{2}}\sum^N_{i=1}\frac{3^{1/4}}
{\sqrt{f(\bn_i)}}
- \gamma_t N
\right],
\end{equation}
where $\gamma_t = 0.783 936 3355$ 
and 
\begin{equation}
f(\bn_i)=\sin^2 \left[\frac{\pi n_{ix}}{2N_s}\right] + 3\sin^2 \left[\frac{\pi n_{iy}}{2N_s}\right].
\end{equation}

By working out the derivatives in Eq.~\eqref{Eqn:C} while using the renormalized distance given in Eq.~\eqref{Eqn:dist_renorm2} and then inserting
the definition of $\bR_i$ for the triangular lattice, i.e., $\bR_i = r_s \sqrt{\pi/(2\sqrt{3})} (n_x, n_y \sqrt{3})^T$, we obtain the following expression for the $\mathbf{C}$ matrix
%
\begin{align}
\notag
&C^{N_s}_{\alpha\beta}(\vec{0})  = -\left(\frac{\pi \sqrt{3}}{2N_s^2}\right)^{3/2} \frac{\delta_{\alpha\beta}}{r_s^3}
\sum_{\bn_i\neq\vec{0}}
\Bigg[
\frac{\cos\left[\frac{\pi n_{i\alpha}}{N_s}\right]}
{{f(\bn_i)^{3/2}}}
 -
\\ &
\frac{\frac34 (\sqrt{3})^{\delta_{\alpha y}}(\sqrt{3})^{\delta_{\beta y}}
\sin^2\left[\frac{\pi n_{i\alpha}}{N_s}\right]}
{{f(\bn_i)^{5/2}}}
\Bigg]
\label{Eqn:C_2D_0}
\\
&C^{N_s}_{\alpha\beta}(\bn_i) = 
\notag
\left(\frac{\pi \sqrt{3}}{2N_s^2}\right)^{3/2} \frac{1}{r_s^3}
\Bigg[
\frac{\delta_{\alpha\beta}\cos\left[\frac{\pi n_{i\alpha}}{N_s}\right]}
{{f(\bn_i)^{3/2}}}
 -
\\ &
\frac{\frac34 (\sqrt{3})^{\delta_{\alpha y}}(\sqrt{3})^{\delta_{\beta y}}
\sin\left[\frac{\pi n_{i\alpha}}{N_s}\right]\sin\left[\frac{\pi n_{i\beta}}{N_s}\right]}
{{f(\bn_i)^{5/2}}}
\Bigg]
\,\,\,\,\,(\bn_ i \neq \vec{0}),
\end{align}
where the summation in Eq.~\eqref{Eqn:C_2D_0} is over all $\bn_i \neq \vec{0}$ inside the CSC.
A similar procedure as described for the 3D bcc lattice in the previous subsection then leads to $\eta_1^{\text{2D}}$.

\subsection{The 1D Wigner Crystal}
As mentioned in the Introduction, in 1D the classical ground-state energy per electron diverges because 
of a non-integrable singularity in the 1D Coulomb potential.
Therefore we will focus here on the calculation of $\eta_1$.
In 1D the $C$ matrix defined in Eq.~\eqref{Eqn:C} is just a scalar that is given by
\begin{align}
C^N(0) &= \frac{\pi^3}{8N^3} \frac{1}{r_s^3}
\sum_{n=1}^{N-1}
\frac{1+\cos^2[\pi n/N]}{{|\sin [\pi n/N]|^3}}
\\
C^N(n) &= - \frac{\pi^3}{8N^3} \frac{1}{r_s^3}
\frac{1+\cos^2[\pi n/N]}{{|\sin [\pi n /N]|^3}} \quad (n\neq 0).
\end{align}
We note that in the limit $N\rightarrow\infty$ we have the following identity
\begin{equation}
\lim_{N\rightarrow\infty}  C^N(0) = \frac{\zeta(3)}{4 r_s^3} 
\end{equation}
in terms of the Riemann $\zeta$ function.
\section{Results}
\label{Sec:results}
\begin{figure}[t]
\includegraphics[width=\columnwidth]{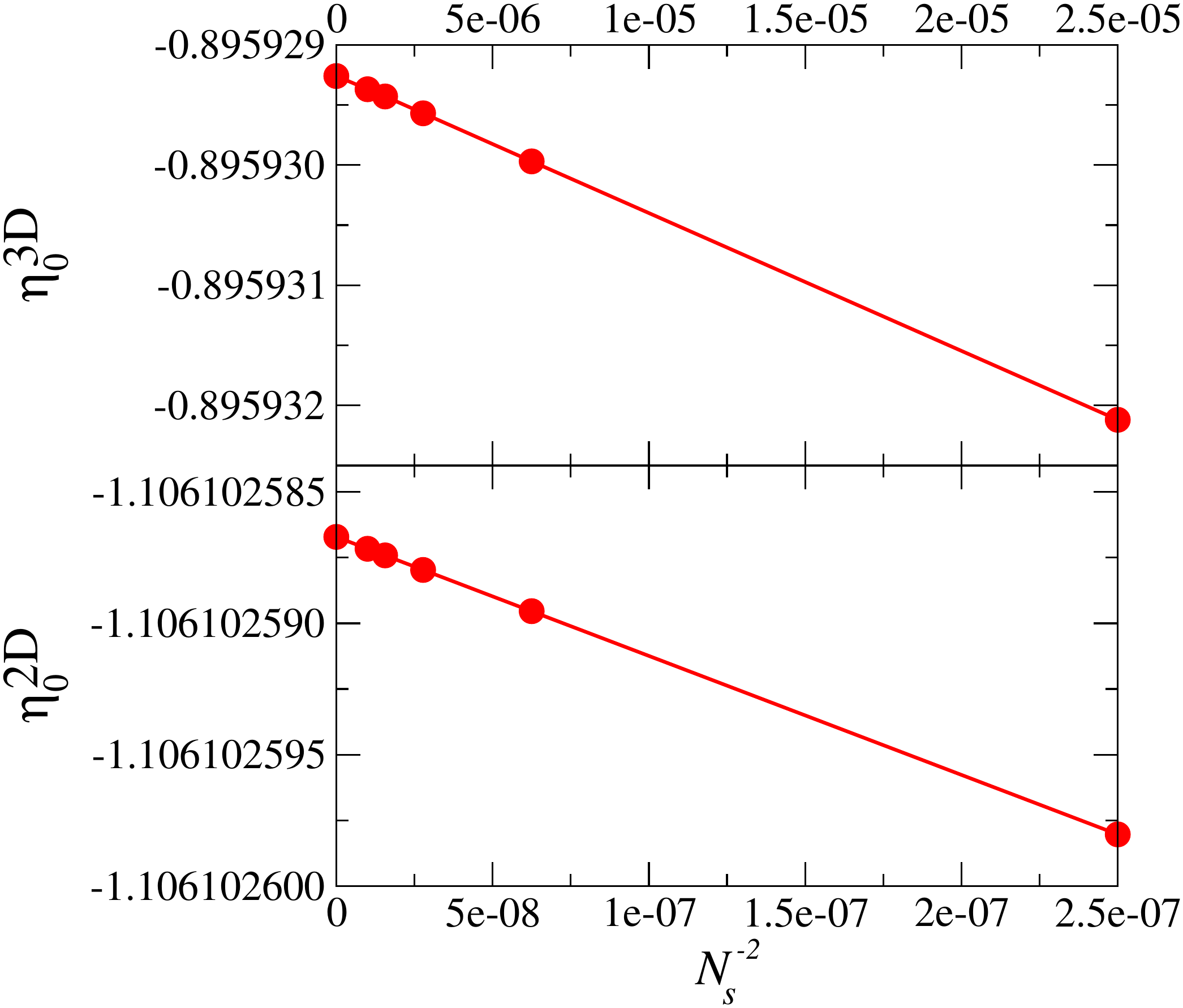}
\caption{The coefficient $\eta_0$ (in Ha) as a function of $N_s^{-2}$ for the 3D bcc lattice and the 2D triangular lattice. 
The dots at $N_s^{-2} = 0$ indicate the
extrapolated values obtained according to Eq.~\eqref{Eqn:extrapol}.}
\label{Fig:eta0}
\end{figure}
\begin{table}[t]
\caption{Numerical values of $\eta_0^{\text{2D}}$ (in Ha) for the square and triangular 2D crystal structures}
\begin{center}
\begin{tabular}{lcc}
\hline
\hline
Lattice &  \multicolumn{2}{c}{$\eta_0^{\text{2D}}$} \\
&  this work &  literature~\cite{Bonsall_1977}    \\
\hline
square \,\,\,\,& \,\,\,\, -1.100 244 420 \,\,\,\, & \,\,\,\, -1.100 244     \\ 
triangle \,\,\,\, & \,\,\,\,  -1.106 102 587 \,\,\,\,  & \,\,\,\, -1.106 103  \\
\hline
\end{tabular}
\end{center}
\label{Table:2D_energies}
\end{table}
We summarize here the results obtained for $\eta_0$ and $\eta_1$ for various lattices in the limit of the infinite systems.
To estimate the coefficients $\eta_0$ and $\eta_1$ for the infinite CSC we extrapolate the coefficients of finite-size CSC with the following power function,
\begin{equation}
 \eta (N_s) = \eta_{\infty} + A N_s^{-2},
 \label{Eqn:extrapol}
\end{equation}
where $A$ and $\eta_{\infty}$ are the fit coefficients.
This power function has also proven to work well for the extrapolation of Madelung constants~\cite{Tavernier_2020}.
\subsection{The classical energy, $\eta_0$}
\begin{table}[t]
\caption{Numerical values of $\eta_0^{\text{3D}}$ (in Ha) for several 3D crystal structures}
\begin{center}
\begin{tabular}{lcc}
\hline
\hline
Lattice &  \multicolumn{2}{c}{$\eta_0^{\text{3D}}$} \\
&  this work &  literature ~\cite{Hasse_1991}  \\
\hline
simple cubic \,\,&\,\, -0.880 059 440 \,\,&\,\, -0.880 059     \\ 
body-centered cubic \,\,&\,\, -0.895 929 255 \,\,&\,\, -0.895 929  \\
face-centered cubic \,\,& \,\, -0.895 873 614 \,\,&\,\, -0.895 874   \\
hexagonal close packed \,\,&\,\, -0.895 838 120 \,\,&\,\, -0.895 838   \\
\hline
\end{tabular}
\end{center}
\label{Table:3D_energies}
\end{table}
\begin{figure}[b]
\includegraphics[width=\columnwidth]{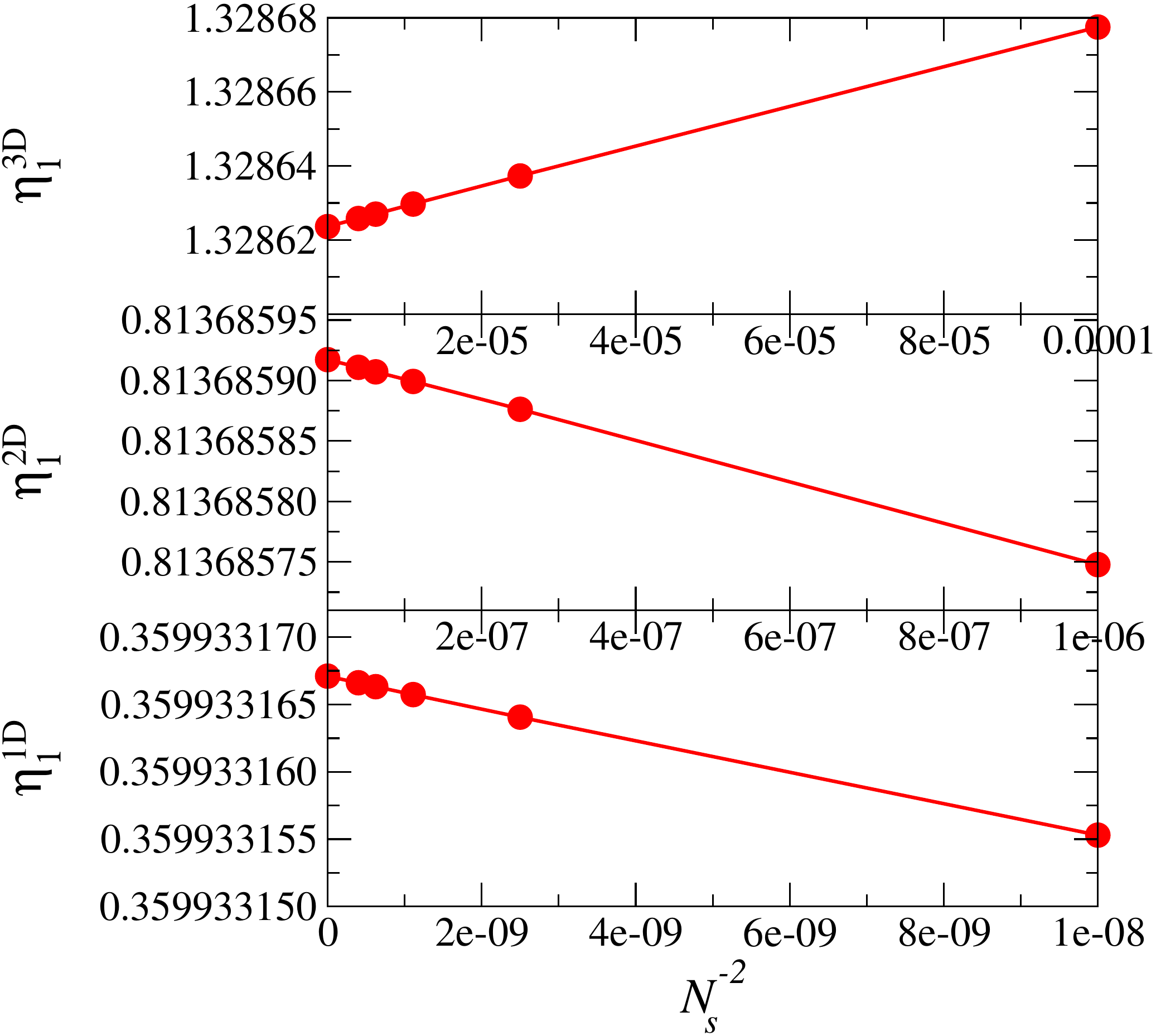}
\caption{The coefficient $\eta_1$ (in Ha) as a function of $N_s^{-2}$ for the 3D bcc lattice, the 2D triangular lattice, and the 1D linear lattice. 
The dots at $N_s^{-2} = 0$ indicate the
extrapolated values obtained according to Eq.~\eqref{Eqn:extrapol}.}
\label{Fig:eta1}
\end{figure}
In Fig.~\ref{Fig:eta0} we report $\eta_0^{\text{3D}}$ and $\eta_0^{\text{2D}}$ as a function of $N_s^{-2}$, for the bcc and triangular lattice respectively. 

We can use the same strategy to find the ground-state energy for any crystal structure.
For the sake of completeness we report in Tables \ref{Table:2D_energies} and \ref{Table:3D_energies} the ground-state energies 
of the most common crystal structures in 2D and 3D, respectively.
As expected we find that in 2D the triangular lattice is lower in energy than the square lattice, while in 3D, it is the bcc lattice that has the lowest energy,
although the difference with the fcc and hcp lattices are small.
Our results are in perfect agreement with the literature values while with our approach we can easily obtain several more digits.
\subsection{The harmonic correction, $\eta_1$}
\begin{table}[t]
\caption{Numerical values of $\eta_1$ (in Ha) for Wigner crystals in 1D, 2D, and 3D}
\begin{center}
\begin{tabular}{lcc}
\hline
\hline
Lattice &  \multicolumn{2}{c}{$\eta_1$} \\
&  this work &  literature   \\
\hline
1D (linear) \,\,&\,\, 0.359 933 \,\,&\,\,  0.359 933 ~\cite{Fogler_2005}    \\ 
2D (triangular) \,\,&\,\, 0.813 686 \,\,&\,\, 0.795 ~\cite{Bonsall_1977}     \\ 
3D (body-centered cubic) \,\,&\,\, 1.328 624  \,\,&\,\, 1.328 62 ~\cite{Carr_1961_Jun}  \\
\hline
\end{tabular}
\end{center}
\label{Table:eta1_extra}
\end{table}
In Fig.~\ref{Fig:eta1} we report $\eta_1$ as a function of $N_s^{-2}$ for 1D, 2D, and 3D.
Again the results are close to linear and we can extrapolate to the infinite-size CSC with the power function of Eq.~\eqref{Eqn:extrapol}. 
We report the extrapolated values in Table \ref{Table:eta1_extra}.
We see that for the 1D and 3D Wigner crystals our results are in agreement with the most accurate values found in the literature.
Instead, for the triangular 2D Wigner crystal our result is significantly different from the literature value.
\section{Conclusions}
\label{Sec:conclusions}
\begin{table}[t]
\caption{Summary of the the coefficients $\eta_0$ and $\eta_1$ (in Ha) obtained in this work.}
\begin{center}
\begin{tabular}{lccc}
\hline
\hline
 & 1D &  2D &  3D   \\
 & \, linear lattice \,& \, triangular lattice \, & \, bcc lattice \,  \\ 
\hline
$\eta_0$ &  - & -1.106 103 & -0.895 929   \\
$\eta_1$ &  0.359 933 & 0.813 686  &   1.328 624  \\
\hline
\end{tabular}
\end{center}
\label{Table:ourwork}
\end{table}
We have presented a simple real-space approach for the calculation of the ground-state energy of Wigner crystals
in one, two and three dimensions. Our approach yields values with high precision for the first two terms in the asymptotic expansion
of the energy per electron of Wigner crystals. Our results are in agreement with the values found in the literature
with the exception of the harmonic correction to the zero-point energy of the 2D triangular Wigner crystal for which we find a significantly larger value than
the one found in the literature. We summarized our results in Table~\ref{Table:ourwork}.
Finally, we note that all our results were obtained with simple computer codes of no more than a few hundred lines, all of which are freely available.~\cite{Wigner_code}
%
%
%

%
\end{document}